\documentclass[twocolumn,aps,floatfix,showpacs,bibnotes,pra,superscriptaddress]{revtex4-1}
\newcommand{\be}{\begin{equation}}
\newcommand{\ee}{\end{equation}}
\newcommand{\bea}{\begin{eqnarray}}
\newcommand{\eea}{\end{eqnarray}}

\usepackage{amsmath}
\usepackage{subfigure}
\usepackage{graphicx}
\usepackage{bm}

\begin{document}

\title{Modeling the transport of interacting matter-waves in disorder by a nonlinear diffusion equation}
\author{E. Lucioni}
\author{L. Tanzi}
\author{C. D'Errico}
\author{M. Moratti}
\author{M. Inguscio}
\author{G. Modugno}
\affiliation{LENS and Dipartimento di Fisica e Astronomia, Universit\'a di Firenze,
  and INO-CNR, 50019 Sesto Fiorentino, Italy}

\begin{abstract}
We model the expansion of an interacting atomic Bose-Einstein condensate in a
disordered lattice with a nonlinear diffusion equation normally used for a variety of classical
systems. We find approximate solutions of the diffusion equation that well reproduce the
experimental observations for both short and asymptotic expansion times. Our study establishes a
connection between the peculiar shape of the expanding density profiles and the microscopic
nonlinear diffusion coefficients.
\end{abstract}

\date{\today}
\maketitle

\section{Introduction} \label{sec:intro}
The interplay of disorder and interactions in quantum systems gives rise to a variety of
interesting phenomena, ranging from glassy quantum phases to non-standard transport properties. In
particular, interactions are known to be able to break the Anderson localization due to disorder,
restoring transport in otherwise insulating systems. The prototypal systems in which such effect
has been studied are one-dimensional disordered potentials, where the expansion dynamics of an
initially localized wavepacket has been extensively investigated both in theory
\cite{Shepe93,Kopidakis08, Pikovsky08, Flach09, Skokos09, Veksler09, Mulansky10, Laptyeva10,
Iomin10, Min, Larcher09} and experiments \cite{Lucioni11}. There is now a general agreement on an
anomalous diffusion process, where the time-exponent of the expansion is smaller than the one found
in linear systems, e.g. $<x^2>\sim t^{\alpha}$, with $\alpha<1$. This subdiffusion is essentially
due to the presence of a local diffusion coefficient that depends on density and therefore
decreases as the wavepacket expands. While the time-evolution of the second moment of the
distribution observed in numerics and experiments agrees with microscopic models, a satisfying
modeling of the evolution of the overall shape of the wavepacket is still missing. A natural
question is whether this expansion process can be modeled with a nonlinear diffusion equation widely employed to describe related transport processes in classical systems \cite{Leibenzon,Ames}, which contains
explicitly a density-dependent diffusion coefficient \cite{Mulansky11,Mulansky12,Laptyeva12,Cherroret}.
Recent numerical studies have indeed established a link between the nonlinear diffusion equation
(NDE) and the asymptotic regime of subdiffusion, employing known self-similar solutions of the NDE
\cite{Mulansky12,Laptyeva12}. However, a comparison with experimental data is not yet possible,
since these solutions do not apply to the limited time interval that is possible to study in
experiments, where normally the shape of the wavepacket changes with time \cite{Lucioni11}.

In this work we study this problem and derive approximate solutions of the NDE for the short-time
regime accessible in experiments. We find a relatively good agreement between the density
distributions measured in the experiment and these solutions, and we identify a time-dependent
exponent that links the evolution of the second moment to the changing shape of the distribution.
While the present experiments lack the necessary spatial resolution, we find that the detailed
study of such shape can give direct evidence of the exponent of nonlinearity of the local diffusion
coefficient.

\section{The disordered, interacting system} \label{sec:system}
In the experiment we employ an ultracold cloud on weakly-interacting bosons in a one-dimensional
optical lattice that mimics a truly disordered potential. More in detail, we realize a
quasiperiodic lattice by perturbing a strong sinusoidal lattice with a secondary one having an
incommensurate spacing. As described in more detail elsewhere \cite{Lucioni11,Modugno09},
non-interacting particles in such potential can be described by the well-known Aubry-Andr\'e
tight-binding (single band) Hamiltonian \cite{Aubry}:
\begin{equation}
H=-J\sum_j (b^\dagger_j b_{j+1}+b^\dagger_{j+1} b_J)+\Delta\sum_j\cos(2\pi\beta j)n_j\,,\label{aa}
\end{equation}
where $b^\dagger_j, b_j$ and $n_j=b^\dagger_jb_j$ are the standard on-site creation, destruction
and number bosonic operators, $J$ is the kinetic (hopping) energy, $\Delta$ is the quasi-disorder
energy and $\beta$ is the ratio of the two lattice spacings. This model is known to show Anderson
localization for $\Delta>2J$, with an essentially energy-independent localization length $\xi\sim
1/\ln(\Delta/2J)$, in units of the main lattice spacing \cite{Roati08}.

\begin{figure}[htbp]
\includegraphics[width=0.95\columnwidth,clip]{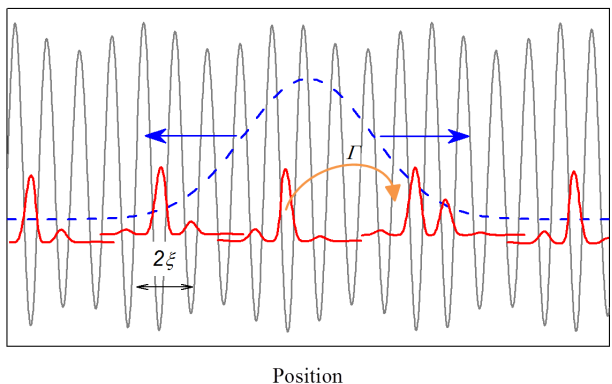}
\caption{(color online) Schematic representation of the disordered, interacting system.
An initial non-interacting wavepacket (blue dashed line) can be decomposed into exponentially-localized
single-particle states (red thick lines) of the quasiperiodic lattice (grey thin line). A weak interaction couples
the states with a density-dependent rate $\Gamma$ and allows an expansion of the wavepacket.
Only states separated by one quasiperiod, 1/($\beta$-1), are shown for clarity. }\label{fig1}
\end{figure}

\begin{figure}[htbp]
\includegraphics[width=0.95\columnwidth,clip]{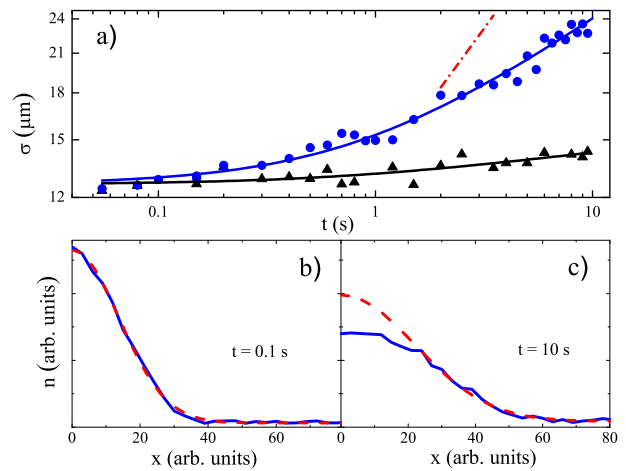}
\caption{(color online) a) Time evolution of the width of the atomic distribution for strong (blue dots)
and very weak (black triangles) interaction.
The continuous lines are fits with Eq.(\ref{eq:m2vst}) and give $\alpha$ exponents of $0.28\pm 0.02$ and
$0.04\pm0.02$ respectively; the dash-dotted line represents the asymptotic slope expected for normal diffusion.
The initial Gaussian distribution (b) evolves into a flat-top distribution at longer times (c). The dashed lines
are Gaussian fits of the tails of the distributions. }\label{fig2}
\end{figure}

In the experiment we can realize this single-particle regime, and also add a controllable repulsive
interaction between the particles, by employing potassium-39 atoms with a magnetically-tunable
Feshbach resonance \cite{Roati07,Derrico07}. In presence of interaction, one needs to introduce an additional term in the
Hamiltonian
\begin{equation}
H_{int}=U\sum_j n_j(n_j-1)\,,\label{Hint}
\end{equation}
where $U$ parameterizes the two-particles interaction energy and $E_{int}\sim U n(x,t)$ represents
the local interaction per particle, where $n(x,t)$ is the time-dependent density distribution. Such interaction can couple distinct single-particle localized states, allowing for macroscopic transport. To probe the transport properties in the experiment, we
initially prepare a low-temperature sample in the combined potential of a quasi-periodic lattice
with $\Delta>2J$ and a harmonic trap, characterized by a Gaussian density distribution $n(x)$. We
then remove suddenly the trap, and let the sample expand along the lattice for a variable time, in
presence of an additional radial confinement. In Fig.\ref{fig1} we show a schematic representation of the experiment. 
As shown in Fig.\ref{fig2}, we essentially observe no expansion if $U$=0, while
for finite $U$ the distribution broadens and changes shape with increasing time. 
The square root of the second moment of $n(x,t)$ increases with a subdiffusive
behavior of the kind
\begin{equation}
\sigma(t)=\sqrt{\langle x^2\rangle}=\sigma_0(1+t/t_0)^\alpha\,,\label{eq:m2vst}
\end{equation}
with a characteristic exponent $\alpha$ in the range 0.2-0.4 for $E_{int}\lesssim
J$ as already discussed in ref.\cite{Lucioni11} (we do not consider the regime of $E_{int}\gg J$, where self-trapping phenomena can complicate the dynamics). While the subdiffusive expansion of the width can be explained with heuristic models of the microscopic
dynamics \cite{Lucioni11,Aleiner}, little or no analysis is available for the evolution of the overall
shape of $n(x,t)$.

Let us start discussing a simple model of the microscopic dynamics which applies to the regime of
weak interactions, where one can still describe the many body-states as superposition of few
single-particle states. A general expectation is that a local diffusion coefficient can be defined
as $D\sim\Gamma\xi^2$, where $\xi$ is the single-particle localization length and $\Gamma$ is the coupling
rate of single-particle states by the interaction. The latter can be evaluated in
perturbation theory as
\begin{equation} \label{eq:Gamma}
\Gamma_{if}= \frac{2\pi}{\hbar}\frac{|\langle i|H_{int}|f\rangle|^2}{|E_i-E_f|}\,.
\end{equation}
where $|i\rangle$ and $|f\rangle$ are two generic initial and final states (they actually
represents quadruplets of single-particle states, because of the form of $H_{int}$) and $|E_i-E_f|$
is their energy separation, which is of the order of $\Delta$. Such coupling is possible only if
$\langle i|H_{int}|f\rangle>|E_i-E_f|$. One can have two different scenarios. If $E_{int}\sim
\Delta$, the dominant couplings are within nearby states, and one finds that $\langle
i|H_{int}|f\rangle$ is essentially $E_{int} I_{if}$, where $I_{if}$ is an overlap integral of the
order unity. This implies that $D\propto n(x,t)^2$ and, since in 1D $n\sim 1/\sigma$, that
$D\propto \sigma^{-2}$. If instead $E_{int}\ll \Delta$, only long-distance couplings are possible,
which tend to decrease with decreasing interaction energy. In this case one must expect $D\propto
n(x,t)^\beta\propto \sigma^{-\beta}$, with $\beta>2$.

By solving the standard diffusion equation
\begin{equation} \label{eq:DE}
\frac{\partial n(x,t)}{\partial t}= D \frac{\partial^2 n(x,t)}{\partial x^2}
\end{equation}
for the width of the distribution, $d\sigma^2(t)/dt=2 D$, one finds a time dependence for
$\sigma(t)$ of the form of Eq.(\ref{eq:m2vst}), with $\alpha=1/2$. If the diffusion coefficient
depends on the width itself as $D\propto\sigma^{-\beta}$, Eq.(\ref{eq:m2vst}) continues to describe
the time evolution of the width but with a time exponent $\alpha=1/(2+\beta)$. These simple
expectations match what has been observed in experiments \cite{Lucioni11} and numerical
calculations (e.g. Ref. \cite{Larcher12} and references therein) for the evolution of $\sigma(t)$.

\section{The nonlinear diffusion equation}
Let us now turn our attention to the evolution of the shape of $n(x,t)$. The idea is to start from
a nonlinear diffusion equation (NDE) of the form
\begin{equation}\label{eq:NDE}
\frac{\partial n(x,t)}{\partial t}= \frac{\partial}{\partial x} \Bigl ( D_0 n^a(x,t) \frac{\partial n(x,t)}{\partial x}\Bigr )\,,
\end{equation}
which takes explicitly into account a density-dependent diffusion coefficient. The NDE is usually
studied in the asymptotic limit, where a self-similar solution exists \cite{Tuck76}:
\begin{equation}\label{eq:asol}
n(x,t)\propto \begin{cases} \Bigl (1-\frac{x^2}{w(t)^2}\Bigr)^{1/a} & \mbox{for } |x| < w(t)
\\0\, & \mbox{for } |x|>w(t) \end{cases}
\end{equation}
The front of the diffusion $w(t)$ has the time dependence $w(t)\propto t^{1/(2+a)}$, and the same
dependence is found for $\sigma(t)$. In our specific problem we expect an exponent $a\geq2$; for
$a=2$ it is exactly an inverted parabola.

This self-similar solution obviously cannot reproduce our short-time expansion, where we clearly
see a changing shape of the distribution. As shown in Fig.2, in the experiment $n(x)$ is initially
Gaussian, while later it develops a flatter top and a relatively faster decay of the tails. This
general behavior is actually consistent with the picture of a density-dependent diffusion
coefficient discussed above, which predicts a larger $D$ at the center of the distribution, and a
reduced one in the tails.

\begin{figure}[htbp]
\includegraphics[width=0.95\columnwidth,clip]{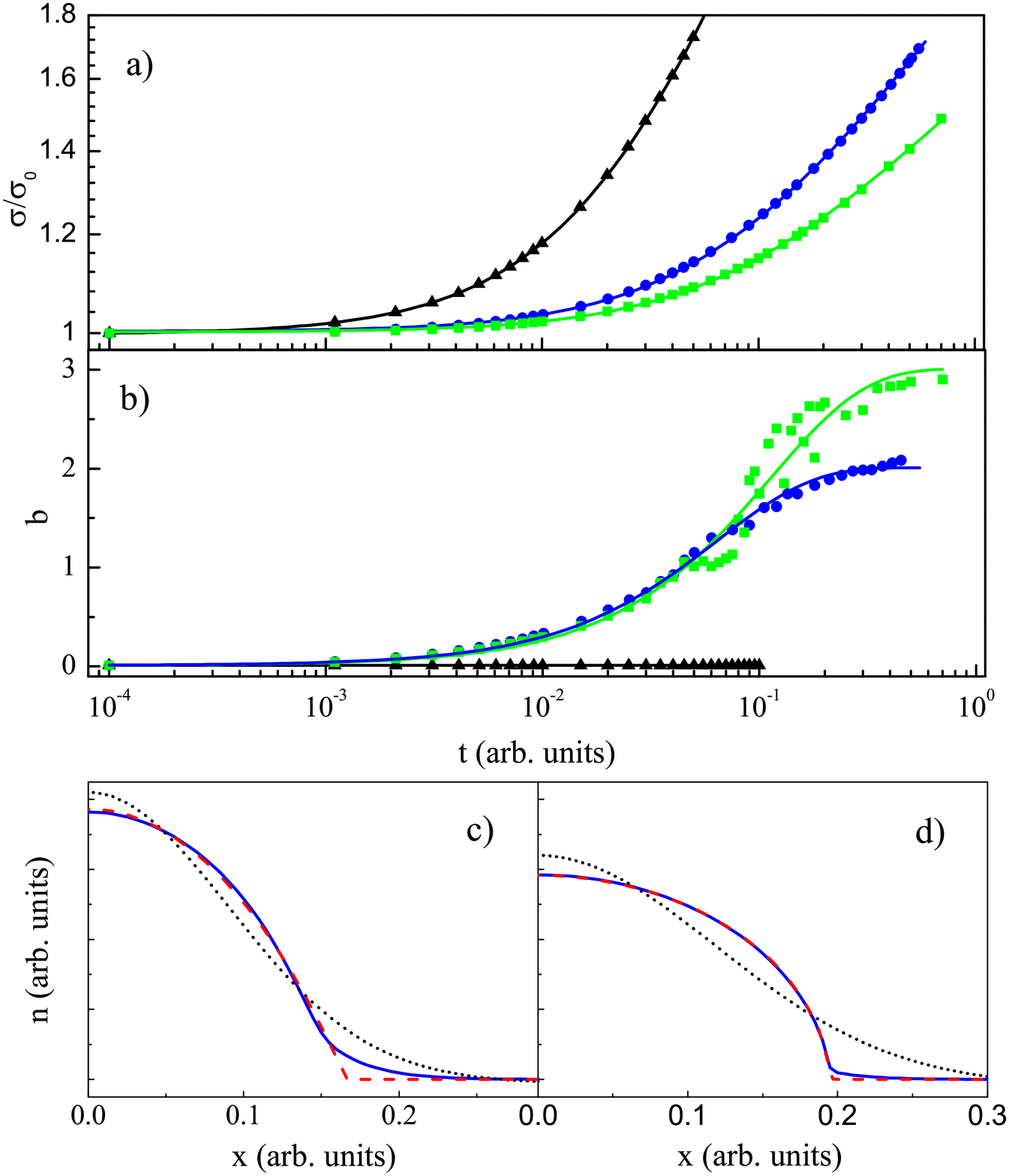}
\caption{(color online) a) Time evolution of the width of the numerical
solution of the NDE (\ref{eq:NDE}) for $a=0$ (black triangles), $a=2$ (blue dots) and $a=3$ (green squares).
Lines are fits with Eq.(\ref{eq:m2vst}) and give $\alpha$ exponents of 0.5, 0.24 and 0.18, respectively.
b) Time evolution of the time-dependent exponent $b$, with same color and symbol scheme as above.
Lines are fits with Eq.(\ref{eq:exp}).
c) Fit of the numerical solution of the NDE for $a=2$ (blue continuous) with Eq.(\ref{eq:approx}) (red dashed)
or a Gaussian (black dotted) for $t$=0.01; d) same as above, for $t$=0.1.}\label{fig3}
\end{figure}

We therefore look for an approximate solution of the NDE which can interpolate between the
initial Gaussian distribution and the asymptotic regime. One can start by noting that a Gaussian
can be obtained as a limit of a slightly different version of Eq.(\ref{eq:asol})
\begin{equation}\label{eq:lim}
e^{-x^2/w^2}=\lim_{b\rightarrow0}\Bigl (1-\frac{bx^2}{w^2}\Bigr)^{1/b}\,.
\end{equation}
The conjecture is then that a solution of the form
\begin{equation}\label{eq:approx}
n(x,t)= \begin{cases}A \Bigl (1-\frac{b(t)x^2}{w(t)^2}\Bigr)^{1/b(t)} & |x| < w(t)/\sqrt{b(t)}
\\0\, & |x|\geq w(t)/\sqrt{b(t)} \end{cases}
\end{equation}
might reproduce the short-time regime of the true solution of the NDE. Here $A= A(b,w)$ is an
appropriate normalization coefficient, and
\begin{equation}\label{eq:exp}
b(t)=a(1-\exp(-t/\tau))
\end{equation}
is a time-dependent exponent. To verify this conjecture, we solve numerically Eq.(\ref{eq:NDE}) for various values of the nonlinear diffusion exponent
$a$, with an initial Gaussian distribution, and we compare the calculated $n(x,t)$ with the approximation above. As summarized in
Fig.\ref{fig3}, we find that this approximation works reasonably well at all times, for values of
the nonlinear diffusion exponent in the range $a=1-3$ (see Appendix B for more details). In
particular, the numerical $n(x)$ is reasonably well reproduced by Eq.(\ref{eq:approx}), besides a
limited deviation of the tails. As we will discuss later, this deviation is not an issue in the
analysis of the experimental data, which has however a limited resolution. There is also a good
agreement of the time evolution of the width with Eq.(\ref{eq:m2vst}), for an exponent $\alpha$ that
is consistent with $\alpha=1/(a+2)$. Finally, $b(t)$ is reasonably well fitted by Eq.(\ref{eq:exp}).

\section{Experimental results}
\begin{figure}[htbp]
\includegraphics[width=0.95\columnwidth,clip]{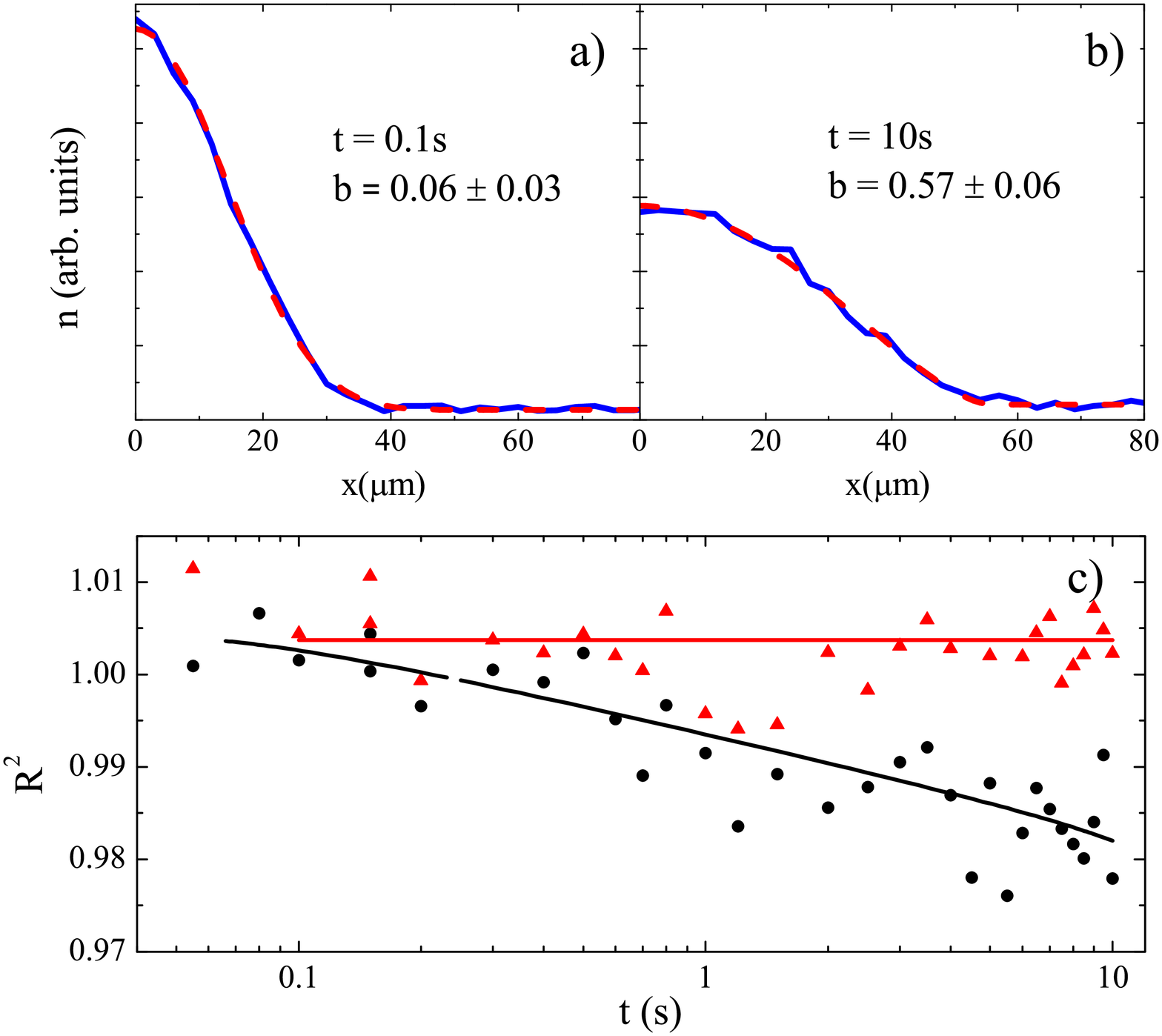}
\caption{(color online) a,b) Fit of the experimental profiles (blue continuous line)
with Eq.(\ref{eq:approx}) (red dashed line). The initial Gaussian distribution
(a) is well fitted with $b \approx 0$, while at longer times (b) the exponent increases.
c) Coefficient of determination $R^2$ for a fit with Eq.(\ref{eq:approx}) (red triangles) and for
a Gaussian fit (black dots). The lines are a guide to the eye.}\label{fig4}
\end{figure}

We can now use Eq.(\ref{eq:approx}) to fit the experimental profiles. Fig.\ref{fig4} shows the
results for a set of experimental data with a mean initial interaction energy $E_{int}=2.3J$ (blue data in Fig.2), with
a rather good agreement. We find that $b$ is close to 0 for short expansion times (Fig.\ref{fig4}a)
and evolves to larger values for increasing expansion times as the flat-top distribution appears
(Fig.\ref{fig4}b). The goodness of the fit for varying time can be evaluated by the coefficient of
determination $R^2$. Fig.\ref{fig4}c shows the evolution of $R^2$ for both a Gaussian fit
and the fit with the approximate solution of the NDE: while the fit with the Gaussian gets worse as the atomic
cloud expands, the fit with Eq.(\ref{eq:approx}) remains constantly good. We note that when the
interaction energy is not strong enough to allow the atomic cloud expansion (black data in
Fig.\ref{fig2}), we cannot appreciate a variation of the the shape of $n(x)$, which is always well
fitted by a Gaussian.

\begin{figure}[htbp]
\includegraphics[width=0.95\columnwidth,clip]{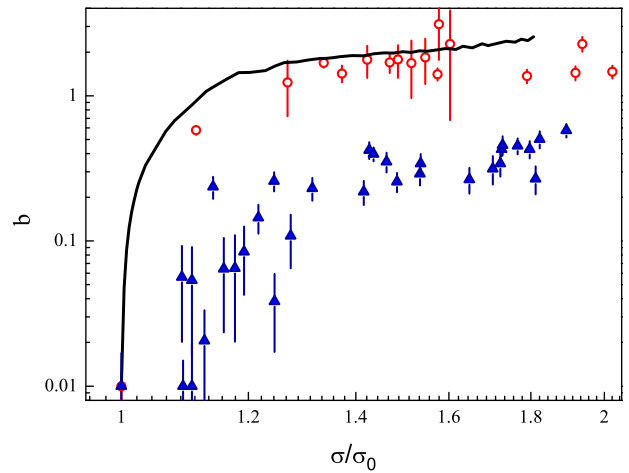}
\caption{(color online) Evolution of the exponent $b$ with the rescaled width
obtained by fitting $n(x,t)$ with the approximate solution of the NDE: experiment (blue triangles)
numerical solution of the NDE with $a=2$ (black line),
numerical solution of the DNLSE (red circles).}
\label{fig5}
\end{figure}

We can now compare the evolution of the exponents $b$ for the experiment and the numerical
solution of the NDE. A direct comparison can be obtained by studying the evolution of $b$ as a
function of the rescaled width $\sigma(t)/\sigma_0$, as shown for example in Fig.\ref{fig5}. This
allows to get rid of the different time and width scales in the experiment and in the simulations.
The experimental data refer to the same set of Fig.\ref{fig4}, for which we measured $\alpha=0.28\pm 0.02$, it is therefore natural to compare it with the
solution of the NDE for $a=2$. One can note a qualitatively similar behavior of theory and
experiment, with an initial $b\approx 0$ that increases towards an asymptotic value. However, the
asymptotic value for the experiment is not the expected one, and the overall evolution is
apparently slower.

We attribute this discrepancy to the finite spatial resolution with which the experimental $n(x)$
is detected. Actually, for our imaging system we have a Gaussian point spread function with a
width of $\sigma_I$=10$\mu$m, which is therefore comparable to the initial width $\sigma_0$. The
expected effect of such finite resolution is indeed a smoothing of the steep decay of the tails of
$n(x)$, and therefore a decrease of the measured exponent $b$. A comparison of the numerical and
experimental profiles in Figs.3-4 confirms this argument.

\begin{figure}[htbp]
\includegraphics[width=0.95\columnwidth,clip]{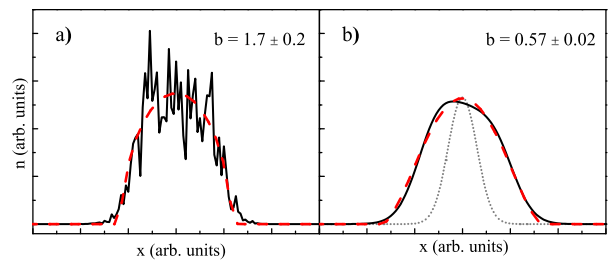}
\caption{(color online) Effect of the finite imaging resolution on the numerical simulations with the DNLSE.
a) $n(x)$ after a large expansion time (black continuous line), in absence of a finite resolution, fitted by Eq.(\ref{eq:approx})
(red dashed line). b) the same $n(x)$ after convolution with Gaussian transfer function (gray dotted line).}
\label{fig6}
\end{figure}

A more quantitative comparison can be made by properly taking the finite resolution into account.
To strengthen our analysis, we also performed a numerical simulation of the expansion by employing
a one-dimensional discrete non-linear Schr\"odinger equation (DNLSE) that is known to reproduce the
evolution of our type of disordered system in the regime of weak interaction. One example of the
numerical $n(t)$ for a long expansion time is shown in Fig.\ref{fig6}. In absence of a finite
spatial resolution (Fig.\ref{fig6}a), one notes rather steep tails that can indeed be fitted with
the approximate solution of the NDE with an exponent $b\approx 2$. Actually, the full evolution of
$b(t)$ for the solution of the DNLSE, which is also reported in Fig.\ref{fig5}, shows a rather good
agreement with the solution of the NDE at all times. When instead the distribution is convolved
with the calculated Gaussian transfer function (Fig.\ref{fig6}b), one can observe a clear smoothing
of the tails, leading to a substantial reduction of $b$.

\begin{figure}[htbp]
\includegraphics[width=0.95\columnwidth,clip]{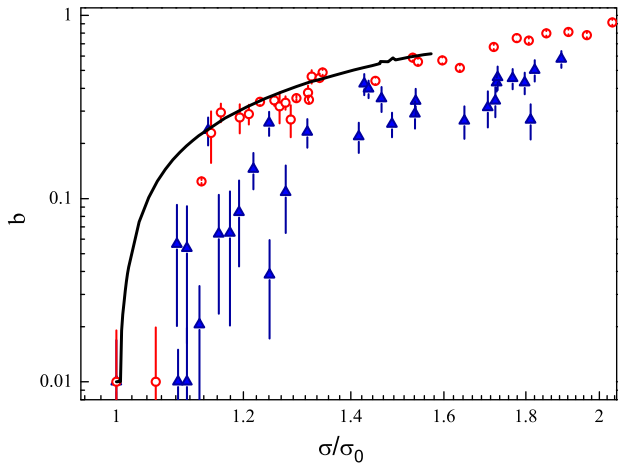}
\caption{(color online) Evolution of the exponent $b$ with the rescaled width
obtained by fitting $n(x,t)$, taking into account the finite spatial resolution: experiment (blue triangles),
numerical solution of the NDE with $a=2$ (black line),
numerical solution of the DNLSE (red circles).}
\label{fig7}
\end{figure}

A direct comparison of the experiment with the NDE can therefore be made only by properly taking
into account such finite resolution also in the numerical solution of the NDE. Fig.\ref{fig7}
compares the exponent $b$ from the experiment with those fitted from the numerical solutions of
both NDE and DNLSE, convolved with the Gaussian point-spread function. The evolution of the NDE
exponent is now slower, and rather close to the experimental one. Clearly, the asymptotic value
$b=a$ can be reached only if the width of the distribution becomes much larger than $\sigma_I$.

The specific set of experimental data we have discussed so far is just one example. We have
actually found similar results for other values of the initial interaction energy in the range
$E_{int}$=0.5-3$J$. For weaker interactions, although a small expansion can be detected in the
experiment (see for example the data in Fig.2), the shape stays essentially Gaussian up to the
longest observation time. We speculate that other effects existing in the experimental setups,
such as a weak time-dependent noise, might be responsible for this observation\cite{Derrico13}.

\section{Conclusions} \label{sec:conclusion}

In conclusion, we have compared the evolution of the density distribution during the subdiffusive
expansion of interacting atoms in disorder with the solution of a general non-linear diffusion
equation of the form of Eq.(\ref{eq:NDE}). In this equation the local diffusion coefficient is
considered to be proportional to the density to some power $a$. To account for the relatively small
timescales available in experiments, we have built an approximate solution of the nonlinear
diffusion equation that provides a sufficiently accurate interpolation between the initial Gaussian
wavepacket and the asymptotic distribution, which is characterized by steep tails. We can see a
qualitative agreement on the evolution of the shape of the distribution, which confirms the
hypothesis of a microscopic density-dependent diffusion coefficient. The finite experimental
spatial resolution did not so far allow us to verify the expected relation between the spatial and
time exponents. This might however be explored in future experiments having state-of-the-art
spatial resolution.

\section{Appendices}
\subsection{Experimental methods and parameters}
The quasiperiodic potential is created by perturbing a deep optical lattice with a weaker lattice
of incommensurate wavelength: $V(x)$=$V_1 \cos^2(k_1 x) + V_2 \cos^2(k_2 x + \phi)$. Here
$k_i$=$2\pi/\lambda_i$ are the wavevectors of the lattices, with $\lambda_1$=1064.4~nm and
$\lambda_2$=859.6~nm, giving a ratio $\beta$=1.238..., $\phi$ is the relative phase between the two lattices. The main lattice fixes the lattice spacing,
$d$=$\lambda_1/2$, and the tunneling energy $J$. The quasi-disorder strength, $\Delta$, scales
linearly with the secondary lattice strength $V_2$ \cite{Modugno09}.

The atomic sample consists in a Bose-Einstein condensate of $^{39}$K atoms in their ground state,
whose $s$-wave scattering length $a_s$ can be adjusted from about zero to large positive values
thanks to a broad magnetic Feshbach resonance \cite{Roati07,Derrico07}. The condensate is initially
produced in a crossed optical trap at $a_s=280 a_0$, and contains about $4\times 10^4$ atoms. A
quasiperiodic lattice with $\Delta\approx3J$ is then slowly added. The radial confinement induced
by the optical trap and the lattice beams is $\omega_r\approx2\pi\times80$~Hz while the axial one
is $\omega_{ax}\approx2\pi\times70$~Hz.

At a given time, $t=0$, the optical trap is suddenly switched off and the atoms are let free to
expand along the lattice, in presence of a radial confinement of $\omega_r\approx2\pi\times50$~Hz
given by the radial profile of the lattice beam. At the same time, $\Delta$ and $a$ are tuned to
their final values within $10$~ms, and kept there for the rest of the evolution. The interaction
parameter is set by the scattering length as
\begin{equation}
U=\frac{2\pi\hbar^2a_s}{m}\int\varphi^4d^3x\,,
\end{equation}
where $\varphi$ is the 3D single-particle Wannier wavefunction. A maximum $E_{int}\approx J$ can be
realized, since an increasing repulsion tends to broaden the system radially, thus reducing its
density.

The atomic density distribution is detected via absorption imaging, and integrated along the radial
direction to obtain the one-dimensional profiles $n(x)$.

\subsection{Approximate solution of the nonlinear diffusion equation}
The complete expression for the approximation to the solutions of the NDE, normalized to unity, is
\begin{equation}
n(x,t)= \frac{b^{3/2}\Gamma(1/b+3/2)}{\sqrt{\pi}w\Gamma(1/b)} \Bigl (1-\frac{bx^2}{w^2}\Bigr)^{1/b}\,,
\label{eq:appendix}
\end{equation}
for $|x|<w/\sqrt{b}$, and zero otherwise. This expression provides an overall better fit of the
numerical solutions of the NDE in all time regimes than either a Gaussian or the asymptotic
solution of the NDE. Fig.8 shows for example the coefficient of determination for the specific case
of $a=2$.

\begin{figure}[htbp]
\includegraphics[width=0.95\columnwidth,clip]{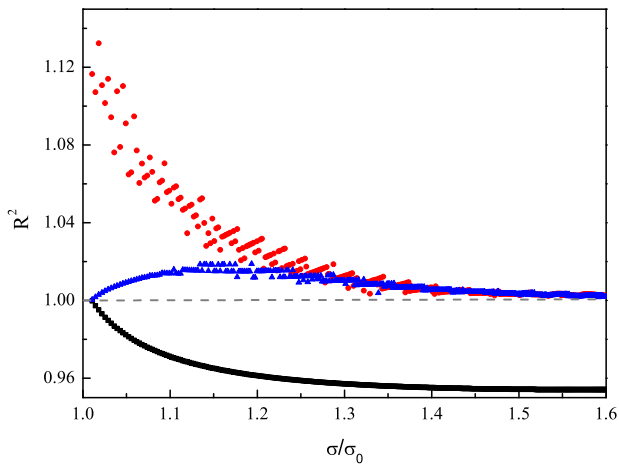}
\caption{(color online) Coefficient of determination $R^2$ of the fit of the numerical solutions
of the NDE with Eq.(\ref{eq:appendix}) (blue triangles), with a Gaussian (black squares) or
with Eq.(\ref{eq:asol}) (red dots).}
\label{fig8}
\end{figure}

\subsection{Numerical simulations}
The DNLSE studied in the numerical simulations reproduces the experimental system in the limit of
negligible population of the radial degrees of freedom:
\begin{eqnarray}\label{}
i\frac{d}{dt}\psi_j(t) = -(\psi_{j+1}(t)+\psi_{j-1}(t))+\nonumber \\
+\Delta/J\sin(2\pi\beta j+\phi)|\psi_j(t)|^2+\gamma|\psi_j|^2\psi_j\,.
\end{eqnarray}
Here $\psi_j(t)$ are the coefficients of the wave function in the Wannier basis, normalized in such
a way that their squared modulus corresponds to the atom density on the j-th site of the lattice.
The mean field interaction strength is given by $\gamma$. The relation between $E_{int}$ and
$\gamma$ is approximately $E_{int}\approx 2 J\gamma/\bar{n}_s$, where $\bar{n}_s$ is the mean
number of sites occupied by the atomic distribution. The initial condition of the simulations is a
Gaussian distribution as in the experiment. For each expansion time, $n(x)$ is obtained by
averaging the profiles resulting from 100 different realizations of the quasiperiodic potential
with randomly varied phase $\phi$ in the range $[0,2\pi]$. The specific results of Figs.5-7 where
obtained with parameters $\gamma=40$ and $\Delta/J=2.5$

\section{Acknowledgments}
This work was motivated by an initial suggestion by Sergej Flach. We acknowledge support by the
European Research Council (grants 203479, QUPOL and 247371, DISQUA), and the Italian Ministry for
Research (PRIN 2009).

\end{document}